\documentclass[printer]{aa}
\usepackage{graphicx}
\newcommand{\bl}{\hbox{BL~Lac} }
\newcommand{\etal}{et al. }

\begin{document}
\title{Multi-band optical micro-variability observations of BL Lacertae}
\author{I.E. Papadakis\inst{1,2} \and P. Boumis \inst{2,3} \and 
V. Samaritakis\inst{2} \and J. Papamastorakis\inst{2,1} }
\offprints{I. E. Papadakis;  e-mail: jhep@physics.uoc.gr}
\institute{IESL, Foundation for Research and Technology-Hellas, P.O.Box
1527, 711 10 Heraklion, Crete, Greece
\and Physics Department, University of Crete, P.O. Box 2208,
   710 03 Heraklion, Greece
\and Present address: Institute of Astronomy \&
Astrophysics, National Observatory of Athens, I. Metaxa \& V. Paulou, P.
Penteli, GR-15236, Athens, Greece} 
\date{Received ?; accepted ?}
\abstract{ We have observed BL Lacertae in the $B$, $R$ and $I$ bands for
2 nights in July, 1999, and 3 nights in July, 2001. The observations
resulted in almost evenly sampled light curves, with an average sampling
interval of $\sim 5$ min. Because of the dense sampling and the
availability of light curves in three bands we are able to study the
intra-night flux and spectral variability of the source in detail. The
source is significantly variable in all bands, showing variations on
different time scales. On average, the variability amplitude increases
from $\sim 5\%$ in the $I$ band, to $\sim 5.5\%$ in the $R$ and $\sim
6.5\%$ in the $B$ band light curves.The rising and decaying time scales
are comparable within each band, but they increase from the $B$, to $R$
and $I$ band light curves. The optical power spectrum shows a red noise
component with a slope of $\sim -2$. Cross-correlation analysis shows that
in most cases the delay between the variations in the $B$ and $I$ band
light curves is less than $\sim \pm 0.4$ hrs. However, the
cross-correlation functions are asymmetric, implying complex delays of the
$I$ band variations with respect to the $B$ band variations. Furthermore,
in one case we find that the $I$ band variations are significantly delayed
(by $\sim 0.2$ hrs) with respect to the $B$ band variations. We also
detect significant spectral variations. We find that the spectrum hardens,
(i.e. it gets flatter) as the flux increases, and the flattest spectral
index corresponds to the maximum $B$ band flux. The rate of the spectral
variations does not remain the same during the observations. Our results
imply that the fast, intra-night variations of the source correspond to
perturbations of different regions in the jet which cause localized
injections of relativistic particles on time scales much shorter that the
average sampling interval of the light curves. The variations are
controlled by the cooling and light crossing time scales, which are
probably comparable.
\keywords{galaxies: active --- galaxies: BL Lacertae objects: general ---
galaxies: BL Lacertae objects: individual: BL Lacertae  ---
galaxies: jets }
}

\titlerunning{Optical micro-variability of BL Lacertae}
\authorrunning{Papadakis \etal}
\maketitle
   
\section{Introduction}
\smallskip

BL Lac objects are one of the most peculiar classes of active galactic
nuclei (AGN). They show high polarization (up to a few percent, as opposed
to less than $\sim 1\%$ for most AGNs) and usually do not exhibit strong
emission or absorption lines in their spectra (but see for example Corbett
\etal 1996, and Vermeulen \etal 1995 for BL Lac itself, and Pian \etal
2002, for PKS 0537-441). They also show continuum variability at all
wavelengths at which they have been observed, from X--rays to radio
wavelengths. In the optical band they show large amplitude, short-time
scale variations. The overall spectral energy distribution of BL Lacs
shows two distinct components in the $\nu-\nu F_{\nu}$ representation. The
first one peaks from mm to the X--rays, while the second component peaks
at GeV--TeV energies (e.g. Fossati \etal 1998). The commonly accepted
scenario assumes that the non-thermal emission from BL Lacs is synchrotron
and inverse-Compton radiation produced by relativistic electrons in a jet
oriented close to the line of sight (e.g. Ghisellini \etal 1998).

BL Lacertae, the object that was used to define this class of AGN, is a
well-studied source that has been observed in the optical band for more
than a century. Large amplitude variations have been observed on both long
and short time scales (Villata \etal 2002, and references therein). \bl
was among the first BL Lac objects which showed large amplitude variations
on time scales of $\sim$ hours (Racine, 1970; Miller \etal, 1989). During
the long, large amplitude 1997 optical outburst the intra-night variations
of the object were studied in detail. Nesci \etal (1998) and Speziali \&
Natali (1998) presented multi-band, micro-variability studies based on
observations that were made in July 1997 and August 1997, respectively. In
both cases, rapid variations were detected in all bands, with their
amplitude increasing towards smaller wavelengths. Matsumoto \etal (1999),
presented $V$ band observations of \bl during August and September, 1997,
and reported the detection of a rapid flux increase of about 0.6 mag
within 40 min. Ghosh \etal (2000), presented $V$ band observations taken
between August and October, 1997. They combined the optical fast,
intra-night variations with simultaneous X--ray and $\gamma-$ray
observations, and analyzed their results within the context of theoretical
models. Clements \& Carini (2001), presented $V$ and $R$ micro-variability
observations that were obtained during the summer of 1997. They observed
significant intra-night variations, which were associated with spectral
variations as well. The spectrum of the source was becoming ``bluer" as it
brightened. Finally, Ravasio \etal (2002) presented simultaneous X--ray
(from {\it Beppo-Sax}) and optical observations of the source during June
and December, 1999. The source has varied continuously in the optical
band. No clear correlation between fast X--ray and optical variability
could be found.

In this work, we present simultaneous, $B$, $R$ and $I$ band monitoring,
intra-night observations of \bl obtained during 1999 and 2001. The
availability of observations in different bands and the almost evenly,
dense sampling pattern of the light curves offer us the opportunity to
study in detail the flux and spectral variations of the source on time
scales of $\sim$ min to hours. The paper is organized as follows. In the
next section we present our observations. In Sect. 3 we estimate the
variability amplitude and compare the observed time scales between the
different band light curves. In Sect. 4 we present the results from a
power spectrum analysis, in Sect. 5 we discuss the spectral variability
properties of the source, and in Sect. 6 we present the cross-correlation
analysis results. A discussion follows in Sect. 7, while a summary of our
work is presented in Sect. 8.

\section{Observations and data reduction} 

BL Lac was observed for 2 nights in 1999 and 3 nights in 2001 from the 1.3
m, f/7.7 Ritchey-Cretien telescope at Skinakas Observatory in Crete,
Greece.  The observations were carried out through the standard Johnson
$B$ and Cousins $R, I$ filters. The CCD used was a $1024 \times 1024$ SITe
chip with a 24 $\mu$m$^{2}$ pixel size (corresponding to
$0^{\prime\prime}.5$ on sky). The exposure time was 180, 60 and 30 sec for
the $B,R $ and $I$ filters, respectively, during the 2001 observing run,
and 120, 60 and 30 sec, respectively, during the 1999 run. In Table 1 we
list the observation dates, and the number of the $B,R$ and $I$ frames
that we obtained each night. During the observations, the seeing was
between $\sim 1^{\prime\prime} - 2^{\prime\prime}$. Standard image
processing (bias subtraction and flat fielding using twilight-sky
exposures) was applied to all frames.

\begin{table}
\begin{center}
\caption{Date of observations, and number of frames (nof) obtained at
each band.}
\begin{tabular}{lccc} \hline
Date & $B$ & $R$ & $I$  \\
     & (nof) & (nof) & (nof) \\
\hline
28/07/99 & 39 & 39 & 39 \\
29/07/99 & 37 & 37 & 37 \\
05/07/01 & 40 & 37 & 38 \\
06/07/01 & 38 & 38 & 38 \\
08/07/01 & 42 & 44 & 42 \\
\hline
\end{tabular}
\end{center}
\end{table}
 
We performed aperture photometry of \bl and of the comparison stars B, C,
H, and K by integrating counts within a circular aperture of radius
$10^{\prime\prime}$ centered on the objects. The photometry of the
comparison stars was taken from Smith \etal (1985) and Fiorucci \& Tosti
(1996) for the $B$ and the $R,I$ bands, respectively. The calibrated
magnitudes of \bl were corrected for reddening and the contribution of the
host galaxy as follows.

For the reddening correction of the nuclear fluxes, we used the
relationship:  $A_{V}=N_{H}/2.23\times10^{21}$ (Ryter 1996), in order to
estimate the $V$-band extinction ($A_{V}$, in magnitudes). We assumed the
column density value of $N_{H}=2.0\times 10^{21}$ cm$^{-2}$, derived from
the X--ray measurements with ROSAT (Urry \etal 1996). This value should be
representative of the nuclear flux absorption, and its use implies
$A_{V}=0.907$. Using the $A_{\lambda}$ versus $\lambda$ relationship of
Cardelli \etal (1989) we found the extinction (in magnitudes) in the $B,
R,$ and $I$ filters: $A_{B}=1.21, A_{R}=0.73$, and $A_{I}=0.54$.

We converted the dereddened magnitudes into flux, and then we corrected
for the contribution of the host galaxy to the measured flux in each band.
According to Scarpa et al. (2000), the $R$ band magnitude of the \bl host
galaxy is $R_{host}=15.55\pm 0.02$. In order to correct $R_{host}$ for
reddening, we used the value of $A_{R}=0.88$ (taken from NED \footnote{The
NASA/IPAC Extragalactic Database (NED) is operated by the Jet Propulsion
Laboratory, California Institute of Technology, under contract with the
National Aeronautics and Space Administration.}). This value is derived
from the Schlegel \etal (1998) maps, and may be more representative of the
Galactic absorption for the host galaxy, which is more extended than the
active nucleus itself. We inferred the $B_{host}$ and $I_{host}$
magnitudes adopting the elliptical galaxy colors (at redshift zero) of
$V-R=0.61, B-V=0.96,$ and $R-I=0.70$ (Fukugita \etal 1995). The resulting
host galaxy fluxes in the $B, R,$ and $I$ bands are 1.40, 4.29, and 6.76
mJy, respectively.  Using the results of Scarpa \etal (2000) and a de
Vaucouleurs $r^{1/4}$ profile, we estimated that the host galaxy
contribution within the circular aperture of radius $10^{\prime\prime}$
centered on \bl should be $70\%$ of the whole galaxy flux. Therefore, the
final $B_{host}, R_{host},$ and $I_{host}$ fluxes that were subtracted
from the observed \bl fluxes were $0.98, 3.0,$ and $4.73$ mJy,
respectively.

\section{The observed light curves} 

\begin{figure}

\resizebox{\hsize}{!}{\includegraphics{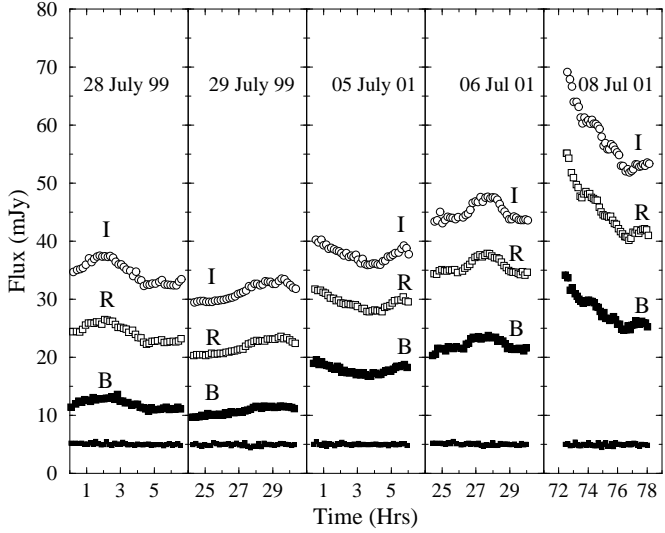}}

\caption[]{$B, R, $ and $I$ band light curves during the 1999
and 2001 observations. Errors are also plotted but are smaller than the
size of the light curve points. Time is measured in hours from 20:00 UT on
July 5, 2001, for the 2001 observations, and from 19:00 UT on July 28,
1999, for the 1999 observations. The small filled squares on the
bottom of the figure show the {\it B} band light curve of the comparison
star B. (For clarity reasons, the light curve is shifted from a mean of
$\sim 7$ mJy to 5 mJy.)}

\end{figure}

\begin{table}
\begin{center}
\caption{The fractional variability amplitude of the light curves at each 
band.}
\begin{tabular}{lccc} \hline
Date & $f_{rms,B}$ & $f_{rms,R}$ & $f_{rms,I}$ \\
     & (\%) & (\%) & (\%) \\
\hline
28/07/99 & 7.2$\pm 0.7$ & 6.2$\pm 0.4$ &
5.7$\pm 0.3$ \\
29/07/99 & 6.9$\pm 0.7$ & 5.5$\pm 0.3$ &
4.8$\pm 0.3$ \\
05/07/01 & 4.2$\pm 0.4$ & 3.9$\pm 0.3$ &
3.5$\pm 0.3$ \\
06/07/01 & 4.3$\pm 0.4$ & 3.6$\pm 0.2$ &
3.6$\pm 0.2$ \\
08/07/01 & 9.1$\pm 0.8$ & 8.8$\pm 0.7$ &
8.5$\pm 0.6$ \\
\hline
\end{tabular}
\end{center} 
\end{table}  

In Fig.~1 we show the final $B,R,$ and $I$ light curves of \bl during our
observations. In the same figure we also show the B band light curve of
the comparison star B. While the light curve of this star (and of the
other 3 comparison stars in all bands) does not show significant
variations, significant intra-night variations can be observed in the \bl
light curves at all bands. During the July $28$ and $29$, 1999
observations we detected two flare-like events at the beginning and at the
end of the observations, respectively. During the July 5 and 8, 2001
observations the flux decreased at the beginning of the night and then,
after $2-4$ hours, it increased again. Finally, a well defined flare which
lasted for $\sim 3$ hours was detected in July 6, 2001. Overall, the
observed light curves show smooth variations which last for a few hours,
with no significant variations on time scales of $\sim$ minutes.

In order to compare the amplitude of the variations that we observe in the
different band light curves, we computed the ``fractional variability
amplitude" ($f_{rms}$) of all the light curves. The fractional variability
amplitude is defined as:
$f_{rms}=(\sigma^{2}-\sigma^{2}_{N})^{1/2}/\bar{x}$, where $\sigma^{2}$ is
the sample variance of the light curve,
$\sigma_{N}^{2}=\sum_{i=1}^{N}err_{i}^{2}/N$ is the variance introduced by
the instrumental noise process ($err_{i}$ is the error associated with
each point, $i$, in the light curve), and $\bar{x}$ is the light curve
mean. The fractional variability amplitude represents the average
amplitude of the observed variations as a percentage of the light curve
mean. The $f_{rms}$ values are listed in Table 2. The errors on the
$f_{rms}$ values were estimated using the bootstrap method of Peterson
\etal (1998). They represent the uncertainty associated with the flux
uncertainties in the individual measurements and the uncertainty
associated with the observational sampling of the light curves (i.e. the
size of the intervals between observations and the length of the light
curves). On average, the amplitude of the observed variations is similar
for both the 1999 and 2001 observations.  Furthermore, the variability
amplitude increases from the $I$ to the $B$ band. The average $f_{rms}$ of
the $B, R,$ and $I$ light curves is $6.3\pm0.3\%, 5.6\pm0.2\%$, and
$5.2\pm0.2\%$, respectively.

\begin{figure}
\resizebox{\hsize}{!}{\includegraphics{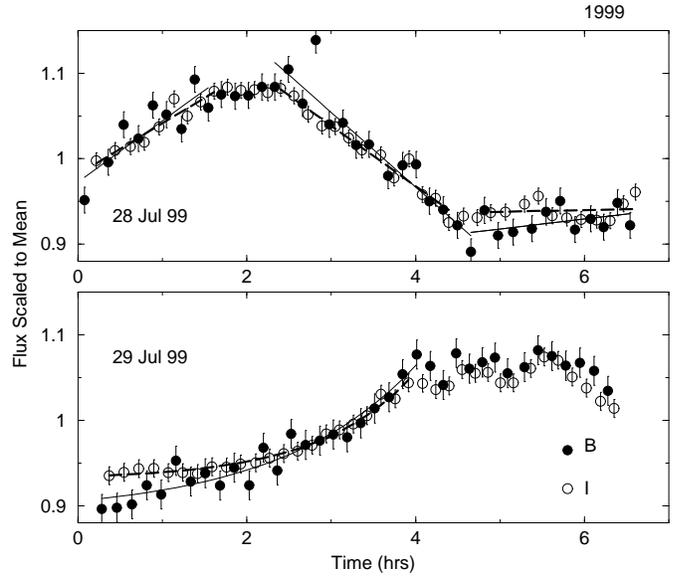}}

\caption[]{$B$ and $I$ band light curves in 1999 (filled and
open circles, respectively). The solid and dashed lines show the best
fitting model to the ``rising" and ``decaying" parts of the $B$ and $I$
light curves, respectively. Time is measured in hours from the beginning
of the observations at each night.}

\end{figure}

\begin{figure}
\resizebox{\hsize}{!}{\includegraphics{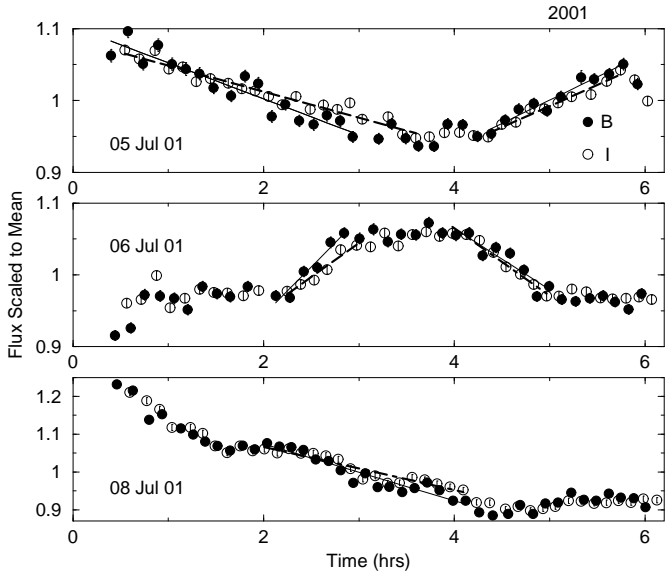}}

\caption[]{Same as Fig.~2, for the 2001 light curves.}

\end{figure}

In order to visualize how well the light curves in the three bands agree
with each other, we normalized them to their mean, and plotted them
together (Fig.~2, and 3 for the 1999 and 2001 light curves, respectively;
for clarity reasons we have plotted only the $B$ and $I$ band light
curves). These plots show that the light curves are well correlated. They
also show that the ``rising"/``decaying" parts of the $B$ band light
curves are steeper than the respective parts in the $I$ band light curves.

In order to compare the variability time scales between the $B, R,$ and
$I$ band light curves, we estimated their ``Flux Variability Rate" ($FVR$)
as follows. First, we identified the rising and decaying parts of the
light curves. Then, we fitted them with a linear model of the form:
$flux(t) = A+B\times t$ (where $t$ is time in hours). The linear model
fits well the rising/decaying phases in all the light curves (as shown by
the solid and dashed lines in Fig.~2 and 3), except for the rising part of
the July 29, 1999 light curves, which was best fitted by an exponential
function of the form: $flux(t)=A_{1}\times \exp (t/\tau)+A_{2}$. Based on
the best model fitting results, we estimated the $FVR$ (in units of
mJy/hrs) for the different phases of each light curve. In effect, the
$FVR$ values show the amount of the flux variation per unit time. The
results for the $B$ and $I$ band light curves are listed in Table 3
(similar model fitting results were also obtained for the $R$ band light
curves as well).

The average $B$ and $I$ band $FVR$ values are: $\bar{FVR}_{B,{\rm
rise}}=7.8\pm0.6$, $\bar{FVR}_{B,{\rm decay}}=-7.4\pm0.5$, and
$\bar{FVR}_{I,{\rm rise}}=6.1\pm0.3$, $\bar{FVR}_{I,{\rm
decay}}=-6.4\pm0.2$ (all values are in units of $\times 10^{-2}$ mJy/hr).  
These results show that the rising $FVR$ values are similar, in absolute
magnitude, to the decaying $FVR$ values, in both the $B$ and $I$ bands.
This result implies that the rising and decaying parts of the light curves
are symmetric. However, when we compare the values between the two bands
we find that the average $FVR_{B}$ values are larger (in absolute
magnitude) than the $FVR_{I}$ band values:  $\bar{FVR}_{B,{\rm
rise}}-\bar{FVR}_{I,{\rm rise}}=1.7\pm0.8$ $\times 10^{-2}$ mJy/hr, and
$\bar{FVR}_{B,{\rm decay}}-\bar{FVR}_{I,{\rm decay}}=-1.0\pm0.5$ $\times
10^{-2}$ mJy/hr.  Therefore, the rising/decaying time scales in the $B$
band are {\it shorter} than in the $I$ band. However this difference is
significant only at the $2\sigma$ level.

Finally, apart from the rising/decaying parts, we can identify at least
two flares with a broad peak during the July 6, 2001 and July 28, 1999
observations. This ``flare plateau" state lasted for $\sim 1$ hr in both
cases and in all light curves.  It is possible that the last part of the
July 29, 1999 observation may also represent the plateau state of yet
another flare. However, we cannot be certain because of the lack of
observations of the decaying phase of the flare.

\begin{table}
\begin{center}
\caption{The ``Flux Variability Rate" ($FVR$, in units of $\times
10^{-2}$
mJy/hr) during the rising and decaying parts of the $B$ and $I$ band
light
curves.}
\begin{tabular}{lcccccc} \hline
Date & $FVR_{B}$ & $FVR_{B}$ & $FVR_{I}$  & $FVR_{I}$
\\
        & (rise) & (decay) & (rise) & (decay) \\
\hline
28/07/99 & $7.1\pm 1.7$ & $-8.7\pm0.9$ & $6.0\pm0.7$ & $-7.1\pm0.5$ \\
29/07/99 & $4.1\pm 0.4$ &$-$ & $3.1\pm0.3$ & $-$ \\
05/07/01 & $6.5\pm 0.5$ & $-5.0\pm0.4$ & $5.9\pm0.3$ & $-3.6\pm0.2$ \\
06/07/01 & $13.4\pm 1.7$ & $-8.5\pm1.5$ & $9.5\pm0.1$ & $ -9.4\pm0.6$ \\
08/07/01 & $-$ & $-7.5\pm0.6$ & $-$ & $-5.7\pm0.5$ \\
\hline
\end{tabular}
\end{center}
\end{table}

\section{Power spectrum analysis}

\begin{figure}
\resizebox{\hsize}{!}{\includegraphics{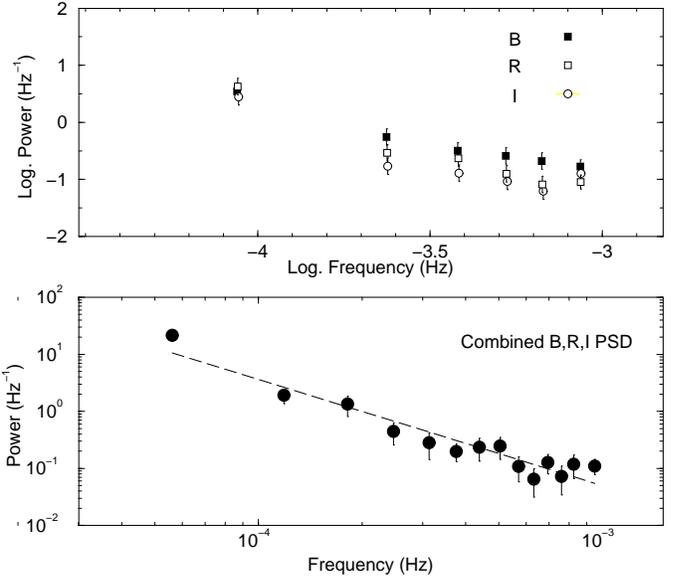}}

\caption[]{Average $B,R,$ and $I$ band power spectra of \bl
(upper panel). In the lower panel, we plot the ``average optical power
spectrum" of the source, with the experimental noise component subtracted.
The dashed line shows the best fitting power law model to the data.}

\end{figure}

In order to quantify the variability seen in the optical light curves of
\bl we estimated their power spectrum as follows. For each individual
light curve, we computed the periodogram as:

\begin{equation}
\hat{I}(\nu_{i})=(\Delta t/N) \{
\sum_{i=1}^{N}[x(t_{i})/\bar{x}-1]e^{-i2\pi\nu_{i}t_{i}} \}^{2},
\end{equation}

where $\bar{x}, \Delta t,$ and $N$ are the mean value, bin size, and
number of points of each light curve, respectively, and
$\nu_{i}=i/(N\Delta t), i=1,2,\ldots, (N/2)-1$ (e.g. Papadakis \&
Lawrence, 1993). The periodogram calculated in this way (i.e. with the
points normalized to the light curve mean) has the units of Hz$^{-1}$.
This normalization is necessary in order to combine the periodograms of
the different light curves (see below). As $\Delta t$ we accepted the
mean interval between the points in each light curve. We note that, due
to the continuous monitoring of the source, the light curves are almost
evenly sampled. The intervals between successive points are almost
equal, and any minor difference cannot affect seriously the estimation
of the power spectrum. Finally, we combined the periodograms of all the
light curves in each band, we sorted them in order of increasing
frequency, we computed their logarithm and grouped them into bins of
size 15 following the method of Papadakis \& Lawrence (1993).

The resulting $B, R,$ and $I$ band power spectra are shown in the upper
plot of Fig.~4 (filled squares, open squares and open circles,
respectively). They all show a similar red noise component, i.e. they
all increase logarithmically towards lower frequencies. The power
spectrum normalization increases slightly from the $I$ to the $B$ band
power spectrum, as expected from the fact that the $B$ band light curves
show the largest $f_{rms}$ values.

Since the three power spectra show a similar shape, we combined the
periodograms of all the light curves in order to estimate an ``average
optical band" power spectrum of the source. Since the Poisson noise
power level is different in each light curve (due to the fact that the
experimental errors are different), we subtracted the expected Poisson
noise level from each periodogram, we sorted them in order of increasing
frequency, and grouped the periodogram estimates into bins of size 20.

The average optical power spectrum of \bl is plotted in the lower plot
of Fig.~4. Using standard $\chi^{2}$ statistics, we fitted the power
spectrum with a power law model of the form $P(\nu)=A\nu^{-a}$. The
model provides a good fit to the data ($\chi^{2}=10.6$ for 12 dof). The
best fitting model is also plotted in Fig.~4 (dashed line). The best
fitting slope value is $a=1.87\pm0.16$ (all errors quoted in the paper
correspond to the $68\%$ confidence region).

\section{Spectral variability}

Since $f_{rms}$ is different in the $B,R,$ and $I$ band light curves, and
there is an indication that the rise/decay time scales may be different as
well, we expect the flux variations to be associated with spectral
variations. In order to investigate this possibility we used the
dereddened light curves, after correction for the host galaxy
contribution, to calculate the two-point spectral indices ($B-R$,
$\alpha_{BR}$, and $B-I$, $\alpha_{RI}$) using the equation,
$\alpha_{12}=\log(F_{1}/F_{2})/\log(\nu_{1}/\nu_{2})$, where $F_{1}$ and
$F_{2}$ are the flux densities at frequencies $\nu_{1}$ and $\nu_{2}$,
respectively.

Figs.~5 and 6 show the $\alpha_{BR}$ and $\alpha_{BI}$ versus the $B$ band
flux plots for the 1999 and 2001 observations, respectively. The
$\chi^{2}$ values for a constant slope show that the variations are
statistically significant. The $\alpha_{BR}$ and $\alpha_{BI}$ variations
are broadly correlated with the source flux in a similar way. Overall, as
the $B$ band flux increases, both $\alpha_{BR}$ and $\alpha_{BI}$ increase
as well. This result implies that the spectrum becomes ``harder" (i.e.
``bluer") as the flux increases. However, the relation between flux and
spectral index is not simple. For example, a linear model of the form
$\alpha = A+B\times flux$, cannot fit well any of the the [$\alpha_{BI},
B$] plots in Fig.~5 and 6 (except for the [$\alpha_{BI}, B$] plot of the
July 29, 1999 observations). Because of this reason, we considered the
spectral slope variations during the different phases of the light curves
(i.e. rising, decaying, plateau) and we present our results below. We
focus our discussion on the $[\alpha_{BI}, B]$ plots mainly, since the
errors on the $\alpha_{BI}$ points are smaller than the errors of the
$\alpha_{BR}$ points.

\begin{figure}
\resizebox{\hsize}{!}{\includegraphics{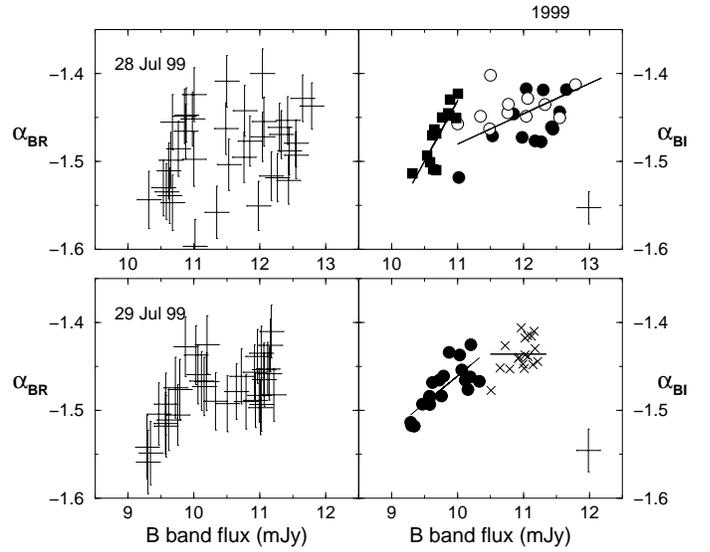}}

\caption[]{$B-R$ and $B-I$ spectral indices vs. the $B$ band source flux
(left and right panels, respectively) for the 1999 observations. The
different symbols for the $B-I$ vs $B$ data in each plot correspond to
different parts of the light curves as explained in the text. The crosses
in the $[\alpha_{BI}, B]$ plots indicate the average error of the points
in the plots.}

\end{figure}

\begin{figure}
\resizebox{\hsize}{!}{\includegraphics{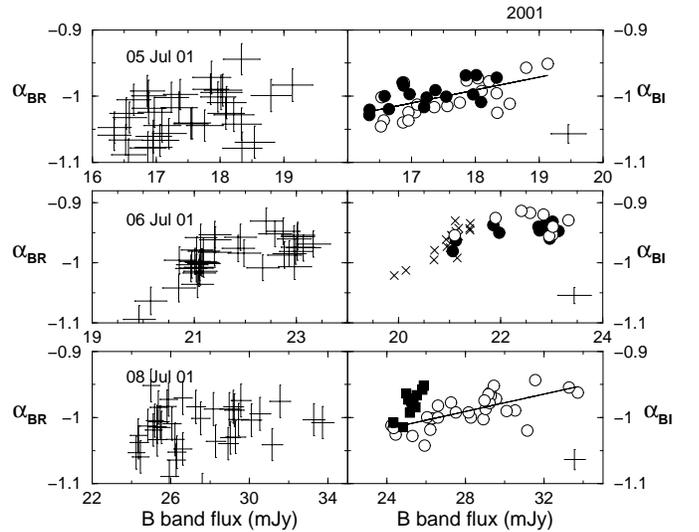}}

\caption[]{ Same as Fig.~5, for the 2001 observations.}

\end{figure}

During the July 28, 1999 observations, the flux increased at the beginning
of the observations, it reached a plateau, then it decreased, and rose
again (at a slower rate) towards the end of the observation (see Fig.~1
and 2). During the rising part of the flare, the spectral slope increased
(solid circles in the upper right plot of Fig.~5), it reached its maximum
value during the plateau state, and then decreased again (open circles in
the same plot) as the source flux decreased. The flux related slope
variations in the two parts of the flare follow the same linear trend
(shown with a solid line in the same plot of Fig.~5). During the last part
of the observation, the moderate flux increase is associated with a marked
spectral slope increase (shown with solid squares). This is caused by the
fact that the $B$ band flux increases faster than the $I$ band flux.
During the rising part of the July 29, 1999 observation the spectral index
increases linearly with the source flux (solid circles in the lower right
plot in Fig.~5; the best fitting linear model is also shown with the solid
line).  It reaches its maximum value during the plateau state (shown with
crosses in the same plot), which lasts until the end of the observation.

A linear relation between spectral slope and source flux is also observed
during the July 5, 2001 observation (upper right plot, Fig.~6). As the
flux decreased during the first part of the observation, the spectral
slope decreased as well (open circles in this plot). Then, when the flux
increased during the second part of the observation, the spectral slope
also increased (solid circles). The linear trend is similar for both the
rise/decay parts of the observation, although the normalization is
different, with the rising part resulting in systematically larger
spectral slope values. A similar behaviour is observed during the July 8,
2001 observation. During the first, long, decaying part of the
observation, the spectral slope decreased (open circles, bottom plot in
Fig.~6). A linear model describes well the overall trend (shown with solid
line in the same plot), but it cannot fit well the data. This is mainly
due to the fact that a few, small amplitude sub-flares can be seen on top
of the long, decreasing phase of the light curves. The moderate flux
increase at the end of the observation is associated with a steep spectral
slope increase (shown with solid squares in the plot). This behaviour is
similar to what was observed at the end of the July 28, 1999 observation.
Finally, during the July 6, 2001 observations, the spectral slope remained
constant ($\alpha_{BI}\sim -0.97$, crosses in the middle plot of Fig.~6)
before and after the flare which started two hours after the beginning of
the observations. The spectral slope increased/decreased together with the
flux in the rise/decay flare phases (shown with solid and open circles,
respectively). The spectral slope reached its maximum value during the
plateau phase.

\section{Cross-correlation analysis}

\begin{figure}
\resizebox{\hsize}{!}{\includegraphics{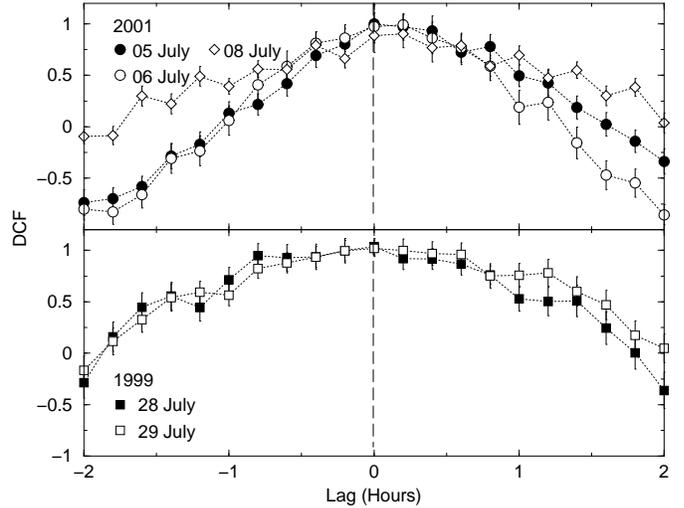}}

\caption[]{Cross-correlation functions between the $B$ band
and $I$ band light curves during the 1999 and 2001 observations (lower and
upper plots, respectively).}

\end{figure}

\begin{table}
\begin{center}
\caption{The cross-correlation analysis results.}
\begin{tabular}{lcc} \hline
Date & $CCF_{max}$ & $k_{max}$ (hrs) \\
\hline
28/07/99 & 1.0 & $-0.07\pm 0.20$ \\
29/07/99 & 1.0 & $+0.10\pm 0.28$ \\
05/07/01 & 0.90 & $+0.23^{+0.19}_{-0.15}$ \\
06/07/01 & 0.96 & $+0.09^{+0.15}_{-0.12}$ \\
08/07/01 & 0.76 & $+0.17^{+0.29}_{-0.26}$ \\
\hline
\end{tabular}
\end{center}
\end{table}

In order to compare the cross-links between the observed variations in the
different bands we estimated the cross-correlation function (CCF) using
the Discrete Correlation Function (DCF) of Edelson \& Krolik (1988).
Fig.~7 shows the DCF between the $B$ and $I$ band light curves during the
1999 and 2001 observations (lower and upper plot, respectively). In these
plots, a positive lag means that the $B$ band leads the $I$ band
variations. In order to quantify the maximum cross-correlation
($CCF_{max}$) and the ``delay" between the light curves (i.e. the time
lag, say $k_{max}$, at which this maximum occurs) we fitted the DCF points
around zero lag with a parabola, and accepted the best fitting values of
the parabola peak as our best estimate of $CCF_{max}$ and $k_{max}$. Our
results are listed in Table 4 (columns 2 and 3 for $CCF_{max}$ and
$k_{max}$, respectively). The uncertainties in the observed lag values
were estimated using the Monte Carlo simulation techniques of Peterson
\etal (1998).

The DCF plots in Fig.~7 and the results on $CCF_{max}$ show that the $B$
and $I$ band light curves are highly correlated, as expected from the
visual good agreement between the different band light curves (e.g.
Fig.~2 and 3). The estimated lag values ($k_{max}$) imply that the delay
between the two band light curves is consistent with zero, within the
errors. Using the cross-correlation results from all the light curves
(except for the July 5, 2001; see below), we estimate an average lag of
$k_{max}=0.07$ hrs, while the $90\%$ confidence limits are $|k_{max}|<
0.42$ hrs (which is $\sim 4$ times the mean sampling rate of the light
curves).

The CCF of the July 5, 2001 light curves, shows an indication that
the $I$ band light curve {\it is} delayed with respect to the $B$ band
light curve. We find a delay of $k_{max}=0.23$ hrs ($\sim$ twice the
mean sampling interval of the light curves) with a $68\%$ confidence
region of [$0.08 - 0.42$ hrs]. In fact, the probability that the $I$
band is delayed with respect to the $B$ band light curve (i.e. the
probability that $k_{max}>0$) is $95.4\%$.

Finally, all cross-correlation functions appear to be asymmetric (except
for the July 28, 1999 CCF). The asymmetry is in the sense that, on time
scales larger than $ 0.5$ hours, the correlation at positive lags is
larger than the correlation at negative lags. In fact, at positive lags,
the $B$ band light curves are better correlated with the $I$ band light
curves than with themselves. This result implies that the variability
components with periods larger than $\sim 0.5$ hours in the $B$ band light
curves lead the respective components in the $I$ band light curves.

\section{Discussion}

Although we cannot estimate systematically time scales from the data
presented in this work, the fact that the duration of the rising/decaying
parts of the light curves is different during our observations suggests
the presence of different variability time scales in \bl. For example, we
observe a flare which lasted for $\sim 3$ hours during the July 6, 2001
observations, while longer flares are detected during the July 28, and
July 29, 1999 observations. Furthermore, the smooth, ``long"-term trends
observed during the July 5, and 8, 2001 observations, are probably parts
of even longer flare-like events. If the difference in time scales
correspond to differences in the size of the emitting source, then perhaps
physically different regions/parts of the jet contribute to the optical
emission of \bl. Even if this is the case, the similarity in the flux and
spectral variability properties suggests that the same physical mechanism
operates in all cases.

The rapid, optical variability properties of \bl are similar to the {\it
X--ray} variability properties of well studied objects like Mkn~421 and
PKS~2155-30. For example, the fractional variability amplitude of the
X--ray light curves in Mkn~421 and PKS~2155-304 varies between $\sim 2\%$
and $\sim 15\%$ (Edelson \etal 2001, Sembay \etal 2002). These values are
comparable to the $f_{rms}$ values listed in Table 2 for \bl. Furthermore,
the X--ray power spectrum of PKS~2155-304 and Mkn~421 shows a red noise
character with a slope of $\sim 2-3$ (Zhang \etal 2002, Kataoka \etal
2001), consistent with the slope of the optical power spectrum of \bl.
Asymmetric CCFs, with no measurable time lags, similar to the CCFs shown
in Fig.~7, have also been observed in PKS~2155-304 and Mkn~421 (Edelson
\etal 2001, Sembay \etal 2002). Finally, there are similarities
between the X--ray spectral variations observed in Mkn~421 and
PKS~2155-304, and the optical band spectral variability of \bl as well.
Fossati \etal (2000) find that, in the case of MKN~421, the spectral slope
at 5 keV decreases (i.e. the spectrum flattens) as the source flux
increases (Figure 4a in their paper). Their result is based on {\it
Beppo-Sax} observations in late April/early May 1997, during which no
individual flares could be observed. Zhang \etal (2002) also find a
similar behaviour between the spectral slope at 0.5 keV and the source flux
in PKS~2155-304, again during observations with no obvious, single flares
(see for example the spectral slope vs flux plot of the ``1997 \#3"
data-set in Figure 11 of their paper). However, clear clockwise and
anti-clockwise ``loop-like" variations of the spectral slope with respect
to the source flux have also been observed in these two X--ray bright BL
Lacs. During the April 1998 {\it Beppo-Sax} observations of MKN~421, the
hard X--ray spectral slope varied with respect to the source flux
following an ``anti-clockwise" loop (Fossati \etal 2000). Similar loops
with a well defined, quasi-circular form have also observed in
PKS~2155-304 (Zhang \etal 2002), mainly during observations dominated by
well sampled, individual flares. In our case, single flares dominate the
observed light curves only during the July 28, 1999 and (mainly) the July
06, 2001 observations. Although in the first case the spectral variations
follow the normal ``flux increase - spectrum flattening" pattern, it is
possible that during the July 06, 2001 observations the variation of the
$\alpha_{BI}$ as a function of the $B$ band flux follows an anti-
clockwise path during the rising and decaying parts of the flare (see
middle panel in Fig.~6). Due to the large errors, we cannot be certain
about the reality of this loop-like variation. Longer and better sampled
observations (with the collaboration of more than one telescope) are
necessary in order to detect and investigate the spectral variations
during individual flares and confirm the existence of loop-like variations
in the optical band ``spectral slope vs flux" plots of BL Lac.

The similarity in the properties of the fast X--ray variations of MKN~421
and PKS~2155-304 and optical variations of BL Lac, may be due to some
physical reason. The multi-frequency spectral energy distribution of \bl
shows a peak at $\sim 10^{14.5}$ Hz while the peak of the synchrotron
emission in the other two objects is located at higher frequencies (e.g.
Sambruna \etal, 1996). Therefore, the optical and X--ray bands correspond
to frequencies that are located around/above the peak of the spectral
energy distribution of \bl on one hand, and Mkn~421 and PKS~2155-30,
on the other. This fact, together with the similarity in their fast
variability properties imply that the intra-night variations in \bl are
probably a direct result of the acceleration/cooling mechanism of
relativistic electrons which represent the highest energy tail of the
synchrotron component, as is generally accepted for the rapid X-ray
variations in the other two sources. In this case, by studying the
optical, rapid variations of \bl we can obtain useful information on the
acceleration ($t_{acc}$) and cooling ($t_{cool}$) time scales of the most
energetic electrons.

Our results suggest that the acceleration process of the energetic
particles does {\it not} dominate the observed variations.  Since the
acceleration time scale should be shorter for lower energy particles, we
should expect the lower energy (i.e the $I$ band) light curves to show
steeper rising phases than the $B$ band light curves. However, we observe
the opposite effect. The $FVR_{B{\rm rise}}$ values are consistently
larger than the $FVR_{R,I{\rm rise}}$ values. Therefore, the $B$ band
light curves rise {\it faster} than $I$ band light curves. Furthermore,
the asymmetry in the CCFs towards positive lags is opposite to what we
would expect if the emission propagates from lower to higher energy. We
conclude that our observations imply that the injection of radiating
electrons in the optical emitting region of \bl is almost instantaneous.

In this case emission should propagate from higher (i.e. the $B$ band)
to lower energy (the $I$ band) and the higher energy should lead the
lower energy photons. This is consistent with the results from the
cross-correlation analysis. Although in most cases we do not detect a
significant delay (except for the July 5, 2001 observations, see below),
the CCF asymmetry towards positive lags implies complex delays between
long period components in the two light curves, in the expected
direction. If the $B$ band variations during the July 05, 2001
observations do lead the $I$ band variations with $k_{max}\sim 0.2$ hrs,
then we can obtain an estimate of the magnetic field strength in the
source. Assuming that the delay is due to synchrotron losses of the high
energy electrons, i.e. $k_{max}=t_{cool}(I)-t_{cool}(B)$, we can use the
following formula (e.g. Chiappetti \etal 1999),

\[ B\delta^{1/3}\sim 
300(\frac{1+z}{\nu_{I}})^{1/3}[\frac{1-(\nu_{I}/\nu_{B})^{1/2}}{k_{max}}]^ 
{2/3}, \] 

to estimate $B$ (in G, when frequencies are given in units of $10^{17}$
Hz). Using $k_{max}=1.7\times 10^{4}$ s$ (=0.2$ hrs), $z=0.0688$, and
$\nu_{B}=0.0068, \nu_{I}=0.0034$ (in units of $10^{17}$ Hz), we find that
$B\delta^{1/3}\sim 1.3$ G, or $B\sim 0.4-0.6$ G, assuming that $10
<\delta<30$ (e.g. Vermeulen \& Cohen, 1994). 

The fact that the rising/decaying time scales of the fully resolved flare
during the July 6, 2001 observations are comparable and roughly similar to
plateau state duration, bears interesting consequences. If $t_{cool}<<R/c$
(where $R$ is the source radius), the energetic particles would reach
equilibrium very fast. The corresponding synchrotron emission should be
switched on and off for a short time and, due to light travel effects,
after a short period (controlled by the short injection time scale,
$t_{inj}$) the observer should see a constant flux, produced by a single
``switched on" slice running across the source (Chiaberge \& Ghisellini,
1999). After the last region of the source is switched on and off, a fast
decline, controlled by $t_{cool}$, should be observed. Therefore, when
both $t_{inj}$ and $t_{cool}$ are much smaller than $R/c$, we expect to
observe a plateau phase which will last much longer than the rise and
decay phase of the flare.  The fact that the July 6, 2001 flare is
symmetric, with rising/decaying/plateau phases which last roughly the same
period, implies that $t_{cool}$ is not much smaller than $R/c$. In this
case, we expect the flux to increase as the observer receives photons from
an increasing volume. At time $t\sim t_{cool}$, the parts of the source
that were seen first will stop emitting, and the flux will remain $\sim$
constant as new parts of the source start to switch on while the parts
closer to the observer are switched off. After $t\sim R/c$, the flux will
start decreasing, as the whole source volume has been activated, and the
front parts of the source keep switching off.

The hypothesis of a jet perturbation with $t_{inj}<<R/c$ and $t_{cool}
\sim R/c$, can also explain the observed spectral variations. As emission
propagates from higher to lower frequencies, it takes time for the higher
energy electrons to cool and start emitting at lower frequencies (i.e. in
the $I$ band). If the increase in the $B$ band emitting volume is faster
than $t_{cool}$, then the higher energy light curves will rise steeper
than the lower energy light curves. As a result, as the $B$ band flux
rises, the observed spectrum will flatten systematically. The
flattest spectral index will correspond to the maximum $B$ band flux, when
the $B$ band emitting region has reached its maximum size, and the volume
of the $I$ band emitting region is still increasing. Furthermore, since
$t_{cool}$ is faster for the $B$ band than the $I$ band emitting
electrons, the decaying phases of the light curves are expected to be
steeper in the $B$ band, as observed. At the same time, the spectrum will
become ``redder" as the $B$ band flux decreases.

\section{Conclusions}

We have observed \bl in three bands, namely $B$, $R$, and $I$, for 2
nights in July, 1999 and 3 nights in July, 2001. On average, each light
curve lasts for $\sim 6$ hours. There are $\sim 40$ points in each of
them, almost evenly spaced, with an average sampling interval of $\sim
0.1$ hrs. Because of the dense sampling and the availability of light
curves in three bands, we were able to study in detail the intra-night
flux and spectral variations of the source. Our results can be
summarized as follows:

1) The source is highly variable in all bands. The variations are smooth,
showing rising/decaying phases which, in some cases, last longer than the
length of each individual light curve. We have also detected 3
``flare-like" events. In particular, during the July 6, 2001 observation,
we observed the whole cycle of a flare which appears symmetric, with a
plateau, and lasted for $\sim$ 3 hours (in all bands). In general, the
rising time scales are comparable to the decaying time scales within each
band. However, these time scales are shorter in the $B$ than the other two
band light curves.

2) The variability amplitude decreases from $\sim 6.5\%$ in the $B$ band,
to $\sim 5.5\%$ and $\sim 5\%$ in the $R$, and $I$ band
light curves, respectively. The average, optical power spectrum of the    
source has a red noise character, with a slope of $\sim -2$ in the
frequency range between [5.5 (hrs$)^{-1} - 15$ (min)$^{-1}$].

3) The light curves in the three bands are well correlated. The variations
occur almost simultaneously in all of them, in the sense that the
delay between the $I$ and $B$ band variations is smaller than $\sim
\pm0.4$ hrs. However, we also find that during the July 5, 2001
observation, there is a $95\%$ probability that the $I$ band light curve
variations are delayed with respect to the $B$ band variations by $\sim
0.2$ hrs. Furthermore, most of the CCFs are significantly asymmetric,
implying complex delays of the $I$ band variations, in all cases.

4) Finally, the source shows significant intra-night spectral slope
variations. These variations are associated with the source flux, in the
sense that the spectrum becomes ``bluer" as the flux increases. The
flattest spectral slope corresponds to the maximum $B$ band flux. The rate
of the spectral slope changes is different for different rising/decaying
parts of the light curves.

Assuming that the variations are caused by perturbations which activate
the jet, the observation of variations with different duration implies
that the perturbations affect different regions of the jet. The fact
that the rising time scales are steeper in the $B$ band light curves and
the CCF asymmetry towards positive lags imply that the injection time
scales are very short (i.e. shorter than our average sampling rate which
is $\sim 3-6$ minutes). These results, together with the observed
spectral variability pattern, imply that the observed variations are
governed by the cooling time scale of the relativistic particles and the
light crossing time scale. The detection of symmetric flares with a
plateau state implies that these time scales are comparable. Finally,
the detection of a soft lag in one of the observations, allows us to
obtain an estimate of the magnetic field strength, $B\sim 0.5$ G.

We believe that our results demonstrate that well sampled, multi-band
optical, {\it intra-night} observations of \bl objects, whose peak of
the emitted power is at mm/IR wavelengths so that the optical emission
corresponds to the emission from the most energetic, synchrotron
emitting electrons in the jet, will offer us important clues on the
acceleration and cooling mechanism of these particles. Since the
injection and cooling times of the particles are very short, light
curves with an average sampling of no more than a few minutes are
necessary to this end.

\vskip 0.4cm

\begin{acknowledgements}
We would like to thank E. Pian, the referee, for helpful comments.  
Skinakas Observatory is a collaborative project of the University of
Crete, the Foundation for Research and Technology-Hellas, and the
Max--Planck--Institut f\"ur extraterrestrische Physik.
\end{acknowledgements}

\end{document}